# Theoretical study on magnetic tunneling junctions with semiconductor barriers CuInSe$_2$ and CuGaSe$_2$ including a detailed analysis of band-resolved transmittances

K. Masuda[1] and Y. Miura[1,2]
[1]Research Center for Magnetic and Spintronic Materials, National Institute for Materials Science (NIMS),
1-2-1 Sengen, Tsukuba 305-0047, Japan
[2]Electrical Engineering and Electronics, Kyoto Institute of Technology, *Kyoto 606-8585, Japan*

We study spin-dependent transport properties in magnetic tunneling junctions (MTJs) with semiconductor barriers, Fe/CuInSe$_2$/Fe(001) and Fe/CuGaSe$_2$/Fe(001). By analyzing their transmittances at zero bias voltage on the basis of the first-principles calculations, we find that spin-dependent coherent tunneling transport of $\Delta_1$ wave functions yields a relatively high magnetoresistance (MR) ratio in both the MTJs. We carry out a detailed analysis of the band-resolved transmittances in both the MTJs and find an absence of the selective transmission of $\Delta_1$ wave functions in some energy regions a few eV away from the Fermi level due to small band gaps in CuInSe$_2$ and CuGaSe$_2$.

## 1. Introduction

Both high magnetoresistance (MR) ratios and low resistance area products (*RA*) are required for MR devices to realize read heads of ultrahigh-density hard disk drives and Gbit-class spin transfer torque magnetic random access memories (STT-MRAMs) [1]. Although magnetic tunneling junctions (MTJs) with insulator MgO barriers have high MR ratios [2,3], these MTJs also have high *RA*. To reduce the *RA*, many studies have used ultrathin MgO barriers (~1nm), by which relatively low *RA* ~ 1 Ωμm$^2$ have been obtained [4)-6]. However, MR ratios also decrease by reducing the thicknesses of the barriers. In addition, ultrathin barriers do not have sufficient controllability in fabrication processes. On the other hand, current-perpendicular-to-plane giant magnetoresistive (CPP-GMR) devices have quite low *RA* because metallic (not insulating) spacers are sandwiched between ferromagnetic electrodes. Although the MR ratio in CPP-GMR devices is increased up to 82% at room temperature (RT) by using Co-based Heusler alloys as electrodes [7], this value is still insufficient for applications. Recently, Kasai *et al.* used a compound semiconductor CuIn$_{0.8}$Ga$_{0.2}$Se$_2$ (CIGS) as a barrier layer of MTJs to achieve high MR ratios and low *RA* [8]. They obtained relatively high MR ratios ~ 40% at RT and low *RA* ranging from 0.3 to 3 Ωμm$^2$ in the MTJs where CIGS is sandwiched between Co$_2$FeGa$_{0.5}$Ge$_{0.5}$ (CFGG) Heusler compounds. Since the barrier thicknesses of these MTJs are twice as those of the above-mentioned MgO-based MTJs with low *RA*, we can expect high controllability in the fabrication processes.

In our previous work [9], we theoretically studied spin-dependent transport properties of two MTJs with different semiconductor barriers, CuInSe$_2$ (CIS) and CuGaSe$_2$ (CGS), to understand the origin of the high MR ratios observed in the CIGS-based MTJs [8]. By analyzing their complex band structures and **k**$_{//}$ dependences of the transmittances, we found that spin-dependent coherent transport of $\Delta_1$ wave functions occurs in both the CIS- and CGS-based MTJs, which can explain the high MR ratios in the CIGS-based MTJs. In our present study, we carry out a more detailed analysis of the CIS- and CGS-based MTJs to understand their transport properties more deeply. We provide the values of calculated conductances in both the MTJs for both the parallel and antiparallel magnetization cases of Fe electrodes. We also analyze the band-resolved transmittances in these MTJs, from which we find an absence of the selective transmission of $\Delta_1$ wave functions in some energy regions a few eV away from the Fermi level due to the small band gaps in the CIS and CGS barriers.

## 2. Method

In this work, we analyzed CIS- and CGS-based MTJs with Fe electrodes: Fe/CIS/Fe(001) and Fe/CGS/Fe(001). Although Heusler alloys were used as electrodes in the experiments on the CIGS-based MTJs, we selected Fe electrodes to understand more simply the transport properties in CIS- and CGS-based MTJs. Since the *a*-axis lengths of CIS (CGS) is close to twice that of bcc Fe, the lattice mismatch between them is considered to be small. We first prepared supercells of Fe/CIS/Fe(001) and Fe/CGS/Fe(001), each of which has two unit cells of CIS (CGS) and one unit cell of Fe on both sides of the CIS (CGS) barrier. The in-plane lattice constant *a* of the supercell was fixed to 0.5782 nm (0.5614 nm), which is an *a*-axis length of the bulk CIS (CGS) [10]. In both the supercells, we optimized all the atomic positions and the distance between the barrier and electrode using the density-functional theory within the generalized

**Table 1** Values of conductances per in-plane unit cell area ($a^2$), $RA_P$ ($RA$ in parallel magnetization states), and MR ratios in (a) Fe/CIS/Fe(001) and (b) Fe/CGS/Fe(001) for different Coulomb interactions $U$. The majority-spin (minority-spin) conductance in the parallel magnetization state is represented as $G_{P,\text{maj.}}$ ($G_{P,\text{min.}}$), and the total-spin conductance in the antiparallel magnetization state is represented as $G_{AP,\text{tot}}$.

(a) Fe/CIS/Fe(001)

| $U$ [eV] | $G_{P,\text{maj.-spin}}$ [$e^2/h$] | $G_{P,\text{min.-spin}}$ [$e^2/h$] | $G_{AP,\text{tot}}$ [$e^2/h$] | $RA_P$ [$\Omega\mu m^2$] | MR ratio [%] |
|---|---|---|---|---|---|
| 0 | $1.923\times 10^{-2}$ | $9.609\times 10^{-3}$ | $1.885\times 10^{-2}$ | 0.299 | 52.9 |
| 5 | $1.662\times 10^{-2}$ | $4.540\times 10^{-3}$ | $1.303\times 10^{-2}$ | 0.408 | 62.3 |
| 10 | $1.427\times 10^{-2}$ | $2.500\times 10^{-3}$ | $9.322\times 10^{-3}$ | 0.515 | 79.9 |

(b) Fe/CGS/Fe(001)

| $U$ [eV] | $G_{P,\text{maj.-spin}}$ [$e^2/h$] | $G_{P,\text{min.-spin}}$ [$e^2/h$] | $G_{AP,\text{tot}}$ [$e^2/h$] | $RA_P$ [$\Omega\mu m^2$] | MR ratio [%] |
|---|---|---|---|---|---|
| 0 | $1.685\times 10^{-2}$ | $5.796\times 10^{-4}$ | $5.854\times 10^{-3}$ | 0.467 | 197.7 |
| 5 | $1.168\times 10^{-2}$ | $2.795\times 10^{-4}$ | $2.983\times 10^{-3}$ | 0.680 | 300.9 |
| 10 | $7.830\times 10^{-3}$ | $1.643\times 10^{-4}$ | $1.673\times 10^{-3}$ | 1.018 | 377.8 |

gradient approximation implemented in the Vienna *ab initio* simulation program (VASP) [11),12)]. In the processes of such structure optimizations, we found that the Se layer is always energetically favored as the termination layer of the barrier in both the CIS- and CGS-based supercells. We studied transport properties of both the MTJs at zero bias voltage with the help of the quantum code ESPRESSO [13)]. This code was applied to the quantum open system composed of the above-mentioned supercell attached to the left and right semi-infinite electrodes of bcc Fe. In the present work, we considered the Coulomb interaction $U$ in the Cu $3d$ states in the barrier. Since the band gap $E_g$ of the barrier increases as $U$ increases, considering $U$ is useful to systematically study the relationship between the band gap and MR ratio. The detailed calculation conditions are the same as our previous study [9)].

Due to the two-dimensional periodicity of our systems in the plane parallel to the layers, the scattering states can be classified by an in-plane wave vector $\mathbf{k}_{//} = (k_x, k_y)$. For each $\mathbf{k}_{//}$ and spin index, we solved scattering equations derived from connection conditions of the wave functions and their derivatives at the boundaries between the supercell and electrodes. Transmittances $T$ were obtained from the transmission amplitudes of the scattering wave functions [14),15)]. We obtained conductances $G$ by substituting the transmittances $T$ into the Landauer formula $G = (e^2/h) \times T$. In this work, we adopted the usual definition of the optimistic MR ratio: MR ratio (%) = $100 \times (G_P - G_{AP})/G_{AP}$, where $G_P$ ($G_{AP}$) is the sum of the majority- and minority-spin transmittances averaged over $\mathbf{k}_{//}$ in the Brillouin zone in the case of parallel (antiparallel) magnetization of Fe electrodes. The values of $RA$ in the parallel magnetization states, $RA_P$, were calculated using the in-plane lattice constants $a$ of the supercells and conductances $G_P$ per in-plane unit cell area ($a^2$) shown in Table 1.

## 3. Results and Discussion

Table 1 shows the values of conductances, $RA_P$, and MR ratios in the CIS- and CGS-based MTJs for different Coulomb interactions $U$. We obtained MR ratios over 50% (190%) for the CIS-based (CGS-based) MTJs. These relatively high MR ratios are due to the spin-dependent coherent transport of $\Delta_1$ wave functions, which was directly confirmed by the $\mathbf{k}_{//}$ dependences of the transmittances [9)]. We also analyzed complex band structures of the CIS and CGS barriers [9)], which also indicated selective transmission of $\Delta_1$ wave functions. The $RA$ values in both the MTJs are much smaller than those of the well-known MgO-based MTJs [9)], which originates from the small band gaps in CIS and CGS barriers.

From Table 1, we see that the MR ratio and $RA$ increase as the Coulomb interaction $U$ increases in both the MTJs. As we mentioned in the previous section, the band gap $E_{g,\text{CIS}}$ ($E_{g,\text{CGS}}$) of CIS (CGS) increases by increasing $U$ from 0 to 10 eV. Actually, we confirmed that $E_{g,\text{CIS}}$ = 0.31, 0.39, and 0.42 eV ($E_{g,\text{CGS}}$ = 0.71, 0.84, and 0.93 eV) for $U$ = 0, 5, and 10 eV, respectively, by analyzing the local density of states of Fe/CIS/Fe(001) and Fe/CGS/Fe(001) [16)]. Thus, it is found that a semiconductor barrier with a larger band gap gives a higher MR ratio and a higher $RA$. This is also supported by the fact that the CGS-based MTJ has a higher MR ratio and a higher $RA$ than the CIS-based one for the same value of $U$ (see Table 1). From the viewpoint of tunneling transport, such an increase in $E_g$ corresponds to an increase in the barrier height $\varphi = E_g - E_F$, where $E_F$ is the Fermi level of the system. Using the above-mentioned band gaps ($E_{g,\text{CIS}}$ and $E_{g,\text{CGS}}$) and the Fermi levels in the Fe/CIS/Fe(001) and Fe/CGS/Fe(001), we obtained the following values of barrier heights: $\varphi_{\text{CIS}}$ = 0.080, 0.100, and 0.111 eV ($\varphi_{\text{CGS}}$ = 0.263, 0.301, and 0.330 eV) for $U$ = 0, 5, and 10 eV, respectively. Note here that the transmittance through the barrier is approximately proportional to exp(-2κd), where $\kappa = (2m\varphi/\hbar^2 - |\mathbf{k}_{//}|^2)^{1/2}$ and $d$ is the barrier thickness [17]. Therefore, increasing $E_g$ (i.e., increasing $\varphi$) in the semiconductor barrier has roughly the same effect as increasing the barrier thickness $d$, both of which sharpen the tunneling feature and enhance the MR ratio.

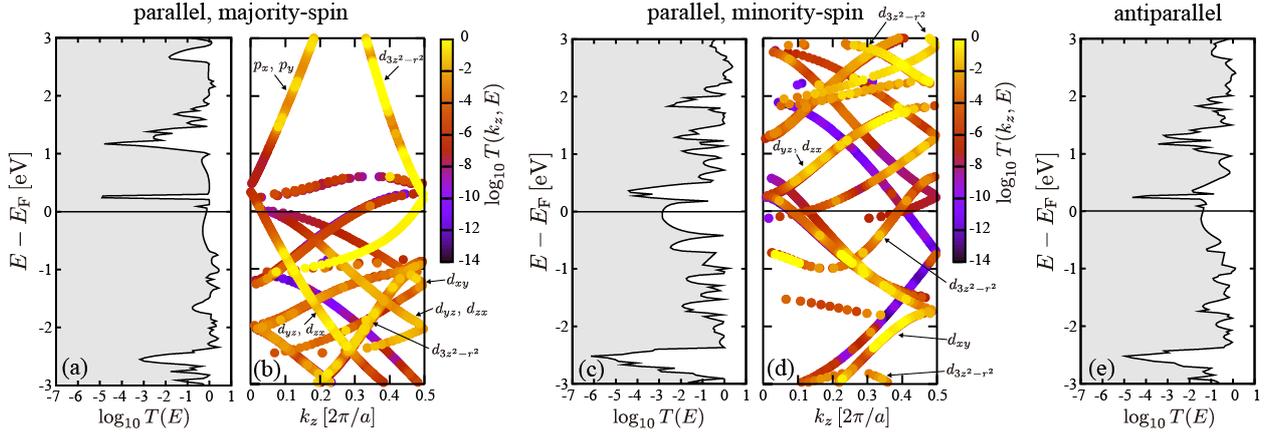

**Fig. 1** Transmittances at $\mathbf{k}_{//}$=(0,0) in Fe/CGS/Fe(001) MTJ with $U$ = 5 eV. (a) The energy dependence of the total transmittance and (b) the band-resolved transmittance in the [001] direction for the majority-spin channel in the parallel magnetization state. (c) and (d) The same as (a) and (b) but for the minority-spin channel in the parallel magnetization state. (e) The energy dependence of the total transmittance in the antiparallel magnetization state.

In order to find characteristic features in the transmittances of the CIS- and CGS-based MTJs, we analyzed energy- and band-resolved transmittances at $\mathbf{k}_{//}$=(0,0) in the CGS-based MTJs [18]. Here, we show the results only for the CGS case, but qualitatively the same results were also obtained in the CIS case. Figures 1(a) and 1(c) show the energy-resolved transmittances for majority-spin and minority-spin channels, respectively, in the parallel magnetization state of the electrodes. At the Fermi level ($E=E_F$), the transmittance in the majority-spin channel is more than one order higher than that in the minority-spin channel. In Fig. 1(e), we also show the energy-resolved transmittance in the antiparallel magnetization state of the electrodes. The transmittance at the Fermi level is one order smaller than the majority-spin transmittance in the parallel magnetization case. From all these behaviors in the transmittances, we can understand naturally the high MR ratio (~300%) in this system (see Table 1). Figures 1(b) and 1(d) show the majority-spin and minority-spin transmittances resolved into each ($k_z$,$E$) contribution on the band structures of the Fe electrode. The energy-gap region is -0.539 eV ≦ $E$-$E_F$ ≦ 0.301 eV for the CGS with $U$ = 5 eV in our calculations [19]. In the majority-spin channel [Fig. 1(b)], the $\Delta_1$ band mainly from $d_{3z^2-r^2}$ state gives high values of transmittance in the energy-gap region. Such a selective transmission of the $\Delta_1$ wave functions is consistent with small imaginary wave vectors in the $\Delta_1$ state obtained in our previous work [see Fig. 2(b) of Ref. 9]. The selective transmission of $\Delta_1$ wave functions also occurs outside of the energy-gap region (0.3 eV ≲ $E$-$E_F$ ≲ 1.1 eV and -1.0 eV ≲ $E$-$E_F$ ≲ -0.54 eV), which is because the conduction and valence bands of CGS in these regions have $\Delta_1$ symmetry. In some other regions (e.g., a region around $E$-$E_F$ = 1.5 eV and that around $E$-$E_F$ = 2.2 eV, etc.), the $\Delta_5$ band mainly from $p_x$ and $p_y$ states gives higher transmittances than the $\Delta_1$ band, which is in contrast to the case of the band-insulator barrier, Fe/MgAl$_2$O$_4$/Fe(001), where the selective transmission of the $\Delta_1$ wave functions persists to high energies [see Fig. 3(b) of Ref. 20 for comparison]. Note again that such energy regions are outside of the band gap of the CGS barrier. Thus, in these regions, tunneling transport is not possible and selective transmission of $\Delta_1$ wave functions do not necessarily occur. We confirmed that the conduction bands of CGS in these regions include $\Delta_5$ component, which is consistent with the high transmittances of $\Delta_5$ states ($p_x$ and $p_y$ states) mentioned above. The similar feature can also be seen in lower-energy regions with $E$-$E_F$ < -1 eV, where the $\Delta_5$ and $\Delta_2$ bands from $d_{yz}$ ($d_{zx}$) and $d_{xy}$ states give relatively high transmittances. This is also supported by the fact that valence bands of CGS in these regions contain $\Delta_5$ and $\Delta_2$ components. Around the Fermi level in the minority-spin channel shown in Fig. 1(d), the $\Delta_5$ and $\Delta_2$ bands give similar values of transmittance as the $d_{3z^2-r^2}$-based $\Delta_1$ band, i.e., clear selective transmission of $\Delta_1$ wave functions does not occur unlike the majority-spin case. Since the in-plane lattice constant of CGS is set to be twice that of the bcc Fe as mentioned above, the original bands of bcc Fe is folded in the $\mathbf{k}_{//}$ Brillouin zone. Usually, in such a case, the folded minority-spin $\Delta_1$ band crossing the Fermi level provides relatively large transmittances and decreases the MR ratio, as discussed in previous studies on Fe/MgAl$_2$O$_4$/Fe(001) [21),22]. However, in our present case, the folded minority-spin $\Delta_1$ band crossing the Fermi level does not give large transmittance compared to the other $\Delta_5$ and $\Delta_2$ bands and therefore does not decrease MR ratio drastically. The configurations of interfacial Se atoms in Fe/CGS/Fe(001) do not have fourfold rotational symmetry along the *c*-axis of the supercell, which is clearly different from the configurations of interfacial O atoms in Fe/MgAl$_2$O$_4$/Fe(001). Thus, in the case of Fe/CGS/Fe(001), although the bands are formally folded in the **k**$_{//}$ Brillouin zone, the folded minority-spin band crossing the Fermi level has a different character from that of Fe/MgAl$_2$O$_4$/Fe(001), which may be the reason for the ineffectiveness of the band-folding effect in Fe/CGS/Fe(001).

The present analysis focused only on the transmittances at zero bias voltage. On the other hand, in actual experiments, a small bias voltage is applied to MTJs to obtain electron transmission. Since the band gaps in the present study are small ($E_{g,CIS} \lesssim 0.42$ eV and $E_{g,CGS} \lesssim 0.93$ eV), such a bias voltage gives a non-negligible spatial gradient in the semiconductor-barrier potentials, which can affect the transport properties in the CIS- and CGS-based MTJs. However, the small values of theoretical band gaps are artifacts due to the generalized gradient approximation used in this study. Although the present band gaps are small, we adopted them as alternatives of experimental band gaps of CIS and CGS ($E_{g,CIS} = 1.0$ eV and $E_{g,CGS} = 1.7$ eV), on which the effect of the spatial gradient from the small bias voltage is not so large.

### 4. Summary

We investigated spin-dependent transport properties of magnetic tunneling junctions with semiconductor barriers, Fe/CIS/Fe(001) and Fe/CGS/Fe(001). We clarified that spin-dependent coherent tunneling transport of $\Delta_1$ wave functions leads to relatively high MR ratios in both the MTJs. From a detailed analysis of the band-resolved transmittances in both the MTJs, we found an absence of the selective transmission of $\Delta_1$ wave functions in some energy regions a few eV away from the Fermi level, which is a characteristic feature of the MTJs with small band gaps in the barrier layers.

**Acknowledgements** The authors would like to thank K. Hono, S. Kasai, and K. Mukaiyama for giving us useful information about experimental results on the CIGS-based MTJs. KM also thanks K. Nawa for valuable discussion on theoretical aspects. This work was in part supported by Grants-in-Aid for Scientific Research (S) (Grant No. 16H06332) and (B) (Grant No. 16H03852) from the Ministry of Education, Culture, Sports, Science and Technology, Japan, by NIMS MI2I, and also by the ImPACT Program of Council for Science, Technology and Innovation, Japan.